\documentclass[amsmath,superscriptaddress,prl,twocolumn,aps]{revtex4}
\usepackage{amsthm,amsfonts,graphicx,verbatim,color}
\newcommand{\be}{\begin{equation}}
\newcommand{\ee}{\end{equation}}
\newcommand{\bea}{\begin{eqnarray}}
\newcommand{\eea}{\end{eqnarray}}

\newcommand{\p}{\partial}

\newcommand{\lp}{\left(}
\newcommand{\rp}{\right)}
\newcommand{\E}{{\cal E}}

\renewcommand{\epsilon}{\varepsilon}
\renewcommand{\vec}[1]{{\bf #1}}

\begin{document}

\title{Giant Spin-Hall Effect induced by Zeeman
Interaction in Graphene}
\author{D. A. Abanin}
\affiliation{Princeton Center for Theoretical Science and Department of Physics, Princeton University, Princeton, NJ 08544}
\author{R. V. Gorbachev}
\affiliation{Manchester Centre for Mesoscience and Nanotechnology, University of Manchester, Manchester M13 9PL, UK}
\author{K. S. Novoselov}
\affiliation{Manchester Centre for Mesoscience and Nanotechnology, University of Manchester, Manchester M13 9PL, UK}
\author{A. K. Geim}
\affiliation{Manchester Centre for Mesoscience and Nanotechnology, University of Manchester, Manchester M13 9PL, UK}
\author{L. S. Levitov}
\affiliation{Department of Physics, Massachusetts Institute of Technology, Cambridge, MA 02139}

\begin{abstract}
We propose a new approach to generate and detect spin currents in graphene, based on a large spin-Hall response arising near the neutrality point in the presence of external magnetic field. Spin currents result from the imbalance of the Hall resistivity for the spin-up and spin-down carriers induced by Zeeman interaction, and do not involve spin-orbit interaction. Large values of the spin-Hall response achievable in moderate magnetic fields produced by on-chip sources, and up to room temperature, make the effect viable for spintronics applications.

\end{abstract}

\maketitle

The spin-Hall effect (SHE) is a transport phenomenon resulting from coupling of spin and charge currents: an electrical current induces a transverse spin current and vice versa\cite{Dyakonov71,Hirsch99}. The SHE offers tools for electrical manipulation of electron spins via striking phenomena such as current-induced
spatial segregation of opposite spins and accumulation of spin at the boundary of current-carrying sample \cite{Wolf2001,Kato04}. 
All SHE mechanisms known to date rely on spin-orbit interaction. The two main varieties of SHE --- intrinsic SHE and extrinsic SHE --- arise due spin-orbit terms in the band Hamiltonian\cite{Sinova04} and spin-dependent
scattering on impurities\cite{Dyakonov71}, respectively.

Single layer graphene 
has emerged recently as an attractive material for spintronics that features long spin diffusion lengths\cite{Tombros07}, gate tunable spin transport\cite{Tombros07,Cho07}, and high-efficiency spin injection\cite{Han10}. However, to realize the full potential of graphene,
several issues must be addressed. First, the measured spin lifetimes 
are orders of magnitude shorter than theoretical predictions\cite{Huertas06,Tombros07,Cho07,Han10,Jozsa09,Pi10} calling for identifying and controlling extrinsic mechanisms of spin scattering\cite{CastroNeto09,Huertas09,Ertler09,Jozsa09,Pi10}. Second, the low intrinsic spin-orbit coupling values\cite{Huertas06,Min06} render the conventional SHE mechanisms ineffective, depriving graphene spintronics of a crucial control knob for spin transport.

\begin{figure}
\includegraphics[width=3.5in]{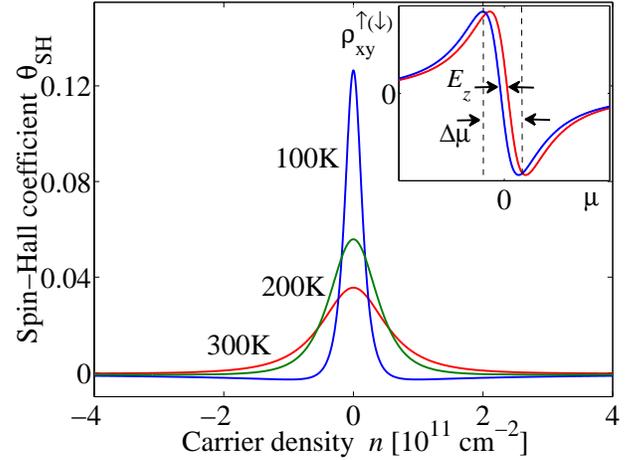} 
\caption[]{Spin-Hall response induced by an external magnetic field in graphene in the absence of spin-orbit coupling. The SHE coefficient $\theta_{\rm SH}$, Eq.(\ref{eq:SHE_DP}), peaks at the Dirac point (DP).
Spin currents at the DP originate from the the imbalance of Hall resistivities for spin up and down due to Zeeman splitting $E_Z$ (inset, red and blue curves). 
Steep behavior of $\rho_{xy}$ leads to large imbalance in the spin-up and spin-down Hall response at chemical potentials $|\mu|\lesssim \Delta\mu$.
Large values $\theta_{\rm SH}$ can be reached already at moderate field strengths and high temperatures, Eq.(\ref{eq:SHE3}).
Parameters used: $B=1\,{\rm T}$, disorder broadening $\gamma=100\,{\rm K}$, electron-hole drag coefficient $\eta={2.3}\hbar$.
}
\label{fig1}
\vspace{-5mm}
\end{figure}

Here we outline a new approach to generate and probe spin currents in graphene, based on a SHE response in the presence of magnetic field that {\it does not rely on spin-orbit interaction}. 
Spin currents are generated by the combined effect of spin and orbital coupling to magnetic field. The Zeeman splitting lifts the up/down spin degeneracy and imbalances the Hall resistivities of the two spin species (see Fig.\ref{fig1} inset). This leads to a net transverse spin current in response to an applied charge current. The resulting SHE response, called below ZSHE for brevity, is an essentially classical effect that offers a robust and efficient way to generate spin currents electrically in a wide range of temperatures and magnetic fields. The ZSHE response is sharply enhanced near the Dirac point (DP). This makes the effect viable for spintronics applications.


The enhancement at the DP, which results from special transport properties of the Dirac fermions, is illustrated in Fig.\,\ref{fig1}. Transport is unipolar at high doping from the DP, dominated by carriers of one type, with $\rho_{xy}$ following the standard quasiclassical expression,
\be\label{eq:rho_xy_quasi}
\rho_{xy}(n)=-\frac{B}{nec}.
\ee
Transport near the DP is bipolar, which produces smearing of the $1/n$ singularity in $\rho_{xy}$ by the effects of two-particle scattering as well as disorder. This leads to a steep linear dependence in $\rho_{xy}(n)$ at the DP (Fig.~\ref{fig1} inset), which is also seen in experiment (Fig.~\ref{fig2}).
The large values of $\partial \rho_{xy}/\partial n$, despite the smallness of the Zeeman splitting, can yield giant ZSHE response.

\begin{figure}
\includegraphics[width=3.3in]{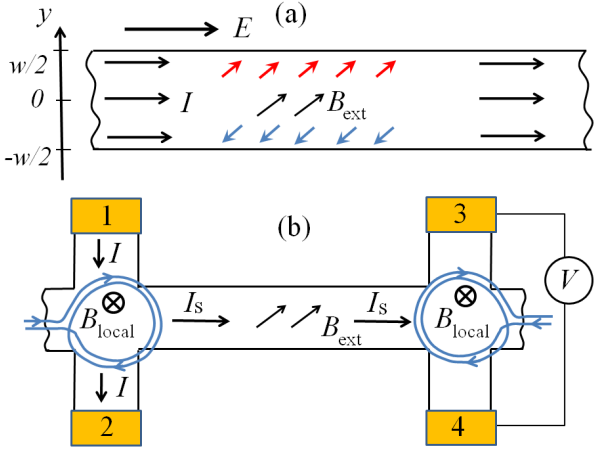} 
\caption[]{(a) Schematic for spin accumulation in the SHE regime. An electric current in a graphene strip drives transverse spin current,
resulting in spin density build-up across the strip, Eq.(\ref{eq:SHE_DP}). (b) Generation and detection of spin current in the H-geometry. Electric current passed through the region of local magnetic field drives spin current along the strip. Voltage generated via inverse SHE is detected using probes 3, 4. Hanle-type oscillation due to spin precession can be induced by external magnetic field applied in-plane.
}
\label{fig5}
\vspace{-5mm}
\end{figure}

The conventional SHE is described by the spin-Hall conductivity which relates transverse spin current and the electric field~\cite{Dyakonov71,Hirsch99}.
To identify the relevant quantity for ZSHE, we consider spin accumulation in a simplified situation when the two spin species are independent, each described by its own conductivity tensor. For a strip of width $w$ carrying uniform current, shown in Fig.\ref{fig5}(a),
the transverse gradients of electrochemical potential for each spin projection are
\be\label{eq:spin_accumulation}
\nabla\lp \phi+\frac{n_\uparrow}{e\nu_\uparrow}\rp =\frac{\rho_{xy}^\uparrow}{\rho_{xx}^\uparrow}\E
,\quad
\nabla\lp \phi+\frac{n_\downarrow}{e\nu_\downarrow}\rp =\frac{\rho_{xy}^\downarrow}{\rho_{xx}^\downarrow}\E
,
\ee
with the up/down spin concentrations $n_{\uparrow(\downarrow)}$ and the densities of states $\nu_{\uparrow(\downarrow)}$.
Ignoring spin relaxation, we estimate spin density at the edge $n_s=n_\uparrow-n_\downarrow$ as
\be\label{eq:SHE_DP}
n_s = \frac{\theta_{\rm SH} we\E}{\nu^{-1}_\uparrow+\nu^{-1}_\downarrow}
,\quad
\theta_{\rm SH}=\frac{\rho_{xy}^{\uparrow}}{\rho_{xx}^{\uparrow}}-\frac{\rho_{xy}^{\downarrow}}{\rho_{xx}^{\downarrow}}
\approx E_Z\frac{\partial }{\partial \mu}\,\frac{\rho_{xy}}{\rho_{xx}}
,
\ee
with $E_Z$ the Zeeman splitting (for full treatment see Appendix A).
Here we used the smallness of $E_Z$ compared to the DP smearing $\Delta\mu$ (see Fig.\,\ref{fig1}) to express $\theta_{\rm SH}$ as a derivative with respect to $\mu$. Our analysis shows that the quantity $\theta_{\rm SH}$ plays a role identical to the ratio of the spin-Hall and ohmic conductivities $\xi_{\rm SH}=2\sigma_{\rm SH}/\sigma_{xx}$ in the conventional SHE. We will thus refer to $\theta_{\rm SH}$ as the SHE coefficient.

For realistic parameter values, Eq.(\ref{eq:SHE_DP}) yields large $\theta_{\rm SH}$ at the peak (see Fig.\ref{fig1}). For $B=1\,{\rm T}$, using disorder strength estimated from mobility in graphene on a BN substrate, $\gamma\approx 100\,{\rm K}$ (see Eq.(\ref{eq:scat_time})), we find  $\theta_{\rm SH}=0.1$. This is more than two orders of magnitude greater than the SHE values in typical spintronics materials with spin-orbit SHE mechanism. Say, we estimate $\xi_{\rm SH}\approx 5\cdot 10^{-4}$ from the spin and charge resistance measured in InGaAs system \cite{Kato04}. The 'giant' values $\theta_{\rm SH}$ are in fact to be expected, since the ZSHE can be viewed as a classical counterpart of the SHE at $k_{\rm B}T< E_Z$ discussed in Refs.\cite{Abanin06,Abanin07} characterized by quantized $\sigma_{\rm SH}=2e^2/h$.

Large $\theta_{\rm SH}$ values result in `giant' spin accumulation. From Eq.(\ref{eq:SHE_DP}), taking $\theta_{\rm SH}=0.1$ and 
the density of states at disorder-broadened DP 
$\nu_{\uparrow(\downarrow)}\approx \sqrt{\Delta n}/\pi\hbar v_0$ 
(with density inhomogeneity $\Delta n\approx 10^{10} {\rm cm^{-2}}$ typical for graphene on BN substrate\cite{Novoselov10}),
and using ${\cal E}=10\, {\rm V/\mu m}$ (which corresponds to maximum current density in graphene~\cite{current_density}), we estimate $n_s$ at the edges of a $2\,{\rm \mu m}$-wide graphene strip:
\be\label{eq:n_s}
n_s \approx 3\cdot 10^{10}\,{\rm cm^{-2}}
.
\ee
Such large densities, which exceed the DP width $\Delta n$, can be easily detected by spin-dependent tunneling. The estimate (\ref{eq:n_s}) is also five orders of magnitude greater than the spin accumulation per atomic layer observed in a three-dimensional GaAs~\cite{Kato04}, $n_s\approx 5\cdot 10^{5} \, {\rm  cm^{-2}}$.

Another attractive feature of the ZSHE is that it can enable {\it local} generation and detection of spin currents. 
Permanent micromagnets can generate fields up to $1\,{\rm T}$ concentrated to regions of size $\sim 0.5\,{\rm \mu m}$~\cite{localB} (fields up to $1.4\,{\rm T}$ are achievable using widely available Neodymium Boron magnets). State-of-the-art microelectromagnets have similar characteristics~\cite{HDDheads}.
In an H-geometry, pictured in Fig.~\ref{fig5}(b), spin currents can be generated on one end of graphene strip and detected on the opposite end. External $B$ field, applied in-plane or at an angle to the graphene sheet, can be used to induce spin precession which will manifest itself in Hanle-type  oscillations of the voltage measured between probes $3$, $4$. This setup can serve as an all-electric probe of spin currents~\cite{Hankiewicz04,Abanin09,Novoselov10}.

To model the dependence of $\theta_{\rm SH}$ on $B$, $T$ and disorder, we employ the quantum kinetic equation approach of Refs.\cite{Kashuba08,Mueller09}. For a spatially uniform system, we have
\be\label{eq:kinetic_equation}
q_{e(h)} \lp \vec{E}+\frac{\vec v}{c}\times \vec{B} \rp \frac{\partial f_{e(h)} (\vec{p})}{\partial \vec{p}}={\rm St}\,[f_{e}(\vec{p}), f_h(\vec{p})],
\ee
where $f_{e(h)}(\vec{p})$ is the distribution function for electrons and holes, and $q_e=-q_h=e$. To describe transport near the DP, it is crucial to account for the contributions of both electrons and holes.
The collision integral describes momentum relaxation due to two-particle collisions and scattering on disorder~\cite{Kashuba08,Mueller09}.

The approach based on Eq.(\ref{eq:kinetic_equation}) is valid in the quasiclassical regime, when particle mean free paths are long compared to wavelength.
This is true when the collision rate is small compared to typical particle energy, which requires weak disorder $\gamma\ll k_{\rm B}T$, where $\gamma$ is defined in Eq.(\ref{eq:scat_time}), and
weak effective fine structure constant $\alpha=e^2/\hbar v_0 \kappa \ll 1$ ($\kappa$ is the dielectric constant).

The kinetic equation (\ref{eq:kinetic_equation}) can be solved analytically in the limit of small $\alpha$~\cite{Kashuba08,Mueller09}.
Rather than pursuing this route, we follow Ref.\cite{Gantmakher78} to obtain transport coefficients from the balance of the net momentum for different groups of carriers, electrons and holes, taken to be moving independently. We use a simple ansatz
\be\label{eq:distributionE}
f_{e(h)}(\vec{p})=\frac{1}{e^{(\epsilon_{\vec{p}}-\vec{p}\vec{a}_{e(h)}- \mu_{e(h)})/k_{\rm B}T}+1}
,\quad
\epsilon_{\vec{p}}=v_0 |\vec{p}|
,
\ee
where $\mu_{e}=-\mu_{h}$ are the chemical potentials of electrons and holes. The quantities $\vec{a}_{e}$ and $\vec{a}_{h}$, which have the dimension of velocity, are introduced to describe a current-carrying state.
This ansatz corresponds to a uniform motion of the electron and hole subsystems, such that the
collision integral for the e-e and h-h processes vanishes (as follows from the explicit form of the collision integral given in Ref.~\cite{Mueller09}). Thus only the e-h collisions contribute to momentum relaxation, resulting in mutual drag between the e and h subsystems.

Eq.(\ref{eq:kinetic_equation}) yields coupled equations for ensemble-averaged velocities and momenta of different groups of carriers (\ref{eq:distributionE}):
\be\label{eq:electron}
q_i \lp {\bf E}+\frac{\vec V_i}{c}  \times {\bf B}\rp=-\frac{\vec P_i}{\tau^{\rm dis}_{i}}-\eta\sum_{i'} n_{i'}   (\vec V_i-{\vec V}_{i'})
,
\ee
where $i$, $i'$ label the e and h subsystems with different spins. The ensemble-averaged scattering times $\tau^{\rm dis}_{i}$, the carrier densities $n_i$, and the electron-hole drag coefficient $\eta$, describing collisions between electrons and holes, are specified below.

The quantities ${\vec V}_i$,  $\vec P_i$ are
proportional to each other, $\vec{P}_i=m_i \vec{V}_i$.
An explicit expression for $m_i$ as a function of $T$, $\mu$ can be found by expanding the distribution functions (\ref{eq:distributionE}) to lowest non-vanishing order in $\vec{a}_{e(h)}$:
\be\label{eq:eff_mass}
m_i = \frac{1}{v_0}
\frac{\int d^2\vec p\, p_x \nabla_{\vec a_x}f_i(\vec p)}{\int d^2\vec p\, \frac{p_x}{p}\nabla_{\vec a_x}f_i(\vec p)}
=\frac{1}{v_0}
\frac{\int d^2\vec p\, p_x^2 g_i(\vec p)}{\int d^2\vec p\, \frac{p_x^2}{p}g_i(\vec p)}
,\quad
\ee
where $g_i(\vec p)=f_i(\vec p)(1-f_i(\vec p))$.
The integrals over $\vec p$, evaluated numerically, give the effective mass as a function of $T$ and $\mu$. At charge neutrality, setting $\mu_{e(h)}=0$, we find $m_T=\frac{9\zeta(3)}{2\zeta(2)} k_{\rm B}T/v_0^2\approx  3.29 k_{\rm B}T/v_0^2$.

The times $\tau^{\rm dis}_{i}$ and carrier densities $n_i$ in
(\ref{eq:electron}) are expressed through the distribution function (\ref{eq:distributionE}) with $\vec a_i=0$:
\be\label{eq:tau_ensem_ave}
\frac1{\tau^{\rm dis}_{i}}=\frac2{n_i} \int\!\!\frac{d^2\vec{p}}{(2\pi)^2} \frac{ f_i(\vec{p})}{\tau^{\rm dis}_{i}(\epsilon_{\vec{p}})}
,\quad
n_i= 2\int\!\!\frac{d^2\vec{p}}{(2\pi)^2}  f_i(\vec{p})
,
\ee
where $\tau^{\rm dis}(\epsilon)$ is the transport scattering time, Eq.(\ref{eq:scat_time}), and the factor of two accounts for valley degeneracy.

We pick the model for disorder scattering to account for the experimentally observed linear dependence of conductivity vs. doping, $\sigma=\mu_* |n|$, where $\mu_*$ is the mobility away from the DP. This is the case for Coulomb impurities or strong point-like defects, such as adatoms or vacancies~\cite{CastroNeto09}. In both cases the scattering time has an approximately linear dependence on particle energy,
\be\label{eq:scat_time}
\tau^{\rm dis}(\epsilon)_{|\epsilon|\gtrsim \gamma}=\hbar |\epsilon|/\gamma^2
, \quad
\gamma=v_0\sqrt{e\hbar/\mu_*}
\ee
where the disorder strength parameter $\gamma$ is expressed through mobility. The value $\mu_*=6\cdot 10^4\,{\rm cm^2/V\cdot s}$ measured in graphene on BN~\cite{Dean10} yields $\gamma\approx 120\,{\rm K}$. Similar values are obtained from the $\rho_{xx}$-based DP width. Taking $\Delta n \approx 10^{10}\, {\rm cm^{-2}}$ \cite{Novoselov10}, we find $\gamma \sim \hbar v_0 \sqrt{\Delta n}\approx 100\, {\rm K}$.

To obtain $\rho^{\uparrow(\downarrow)}_{xy}$, we solve Eq.(\ref{eq:electron}), accounting only for the drag between electrons and holes of the same spin. It can be shown
that including the drag between species of opposite spin does not change the overall behavior of the transport coefficients and SHE (see Appendix C). Eq.(\ref{eq:electron}) can be conveniently analyzed using complex-valued quantities
$P_x+iP_y$, $V_x+iV_y$, giving complex resistivity
\be\label{eq:rho+i*rho}
\rho^{\uparrow(\downarrow)}_{xx}+i\rho^{\uparrow(\downarrow)}_{xy}=\frac1{e^2}\frac{\tilde\gamma_e\tilde\gamma_h+\eta\frac{n_e}{m_h}\tilde\gamma_e + \eta\frac{n_h}{m_e}\tilde\gamma_h}{\frac{n_e}{m_e}\tilde\gamma_h +\frac{n_h}{m_h}\tilde\gamma_e + \eta\frac{(n_e-n_h)^2}{m_e m_h}}
.
\ee
Here $\tilde\gamma_{i}=\frac{1}{\tau^{\rm dis}_{i}} -i \Omega_i$, with $\Omega_{i}=q_i B/m_ic$ the cyclotron frequency.

As a sanity check, we consider the behavior at charge neutrality. Setting $n_e=n_h$, $m_e=m_h$, etc., gives $\rho_{xx}$ which is a sum of the Drude-Lorentz resistivity and the electron-hole drag contribution analyzed in Refs.\cite{Kashuba08,Mueller09},
\be\label{eq:rho_xx_DP}
\rho^{\uparrow(\downarrow)}_{xx}=\frac{m_T}{2 n_T e^2\tau}\lp 1+\tau^2\Omega^2\rp +\frac{\eta}{e^2}
,\quad
n_T=\frac{\pi}{12}\frac{k_{\rm B}^2T^2}{\hbar^2 v_0^2}
,
\ee
and $\rho^{\uparrow(\downarrow)}_{xy}=0$.
Here $n_T$ is the density of thermally activated electrons (holes) at the DP, having fixed spin projection.
Disorder scattering (first term) dominates at low temperatures $T\lesssim T_*=\gamma\sqrt{\hbar/\eta}$ (at $B=0$), while electron-hole drag (last term) dominates at $T\gtrsim T_*$.

The value for the electron-hole drag coefficient $\eta$ can be obtained by matching the last term in Eq.(\ref{eq:rho_xx_DP})), divided by $2$ to account for spin,
to the analytic result $\rho_{xx}\approx 8.4 \hbar \alpha^2/e^2$ \cite{Kashuba08,Mueller09}.  We evaluate $\alpha$ using the effective dielectric constant $\kappa=\frac{\epsilon_0+1}{2}+\frac{\pi}{2}\frac{e^2}{\hbar v_0}\approx 6$, which accounts for screening by substrate and for intrinsic screening in the RPA approximation.
Taking $\epsilon_0\approx 4$ for BN substrate~\cite{Dean10}, yields $\alpha\approx 0.37$, giving $\eta\approx 2.3\hbar$.

\begin{figure}
\includegraphics[width=3.1in]{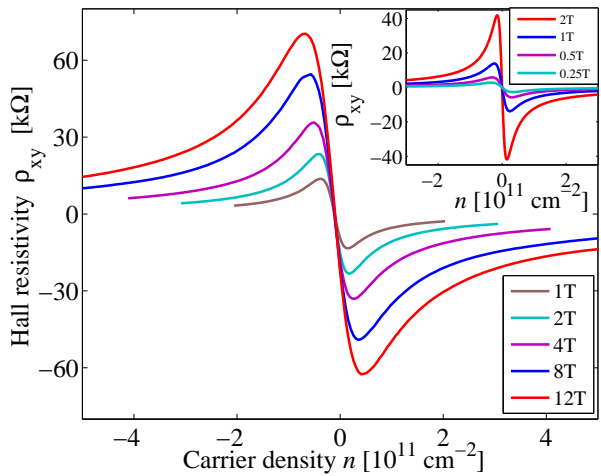} 
\caption[]{Measured $\rho_{xy}(n)$ for a high-mobility graphene sample on BN substrate at $T=250\, {\rm K}$. The dependence follows the quasiclassical formula (\ref{eq:rho_xy_quasi}) away from the DP, and is linear with a steep slope near the DP.
Inset: Results for $\rho_{xy} (n)$ obtained
from the two-carrier model, Eqs.(\ref{eq:electron}),(\ref{eq:rho+i*rho}),
for disorder strength $\gamma=180\, {\rm K}$ found by fitting the min/max distance in measured $\rho_{xy}$ for $B=1\,{\rm T}$.
Other parameters: $\eta={{2.3}}\hbar$, $T=250\, {\rm K}$.
}
\label{fig2}
\vspace{-6mm}
\end{figure}

The dependence of transport coefficients on $T$, $B$ and carrier density $n$, predicted from Eq.(\ref{eq:rho+i*rho}), can be directly compared to experiment. Fig.~\ref{fig2} shows $\rho_{xy}(n)$ measured in graphene on BN, on samples similar to those described in Ref.\cite{Novoselov10}. The modeled $\rho_{xy}(n)$ captures the main features of the data: the $1/n$ dependence at large $n$ and a steep linear dependence near the DP. The linear region
broadens with temperature at $T\gtrsim \gamma$. The peak in $\rho_{xx} (n)$  features similar thermal broadening (see Appendix C).
The SHE coefficient, found from Eq.(\ref{eq:SHE_DP}), is plotted in Fig.\ref{fig1}.



We now explore the behavior of transport coefficients near the DP, making estimates separately for $T\gtrsim T_*$ and $T\lesssim T_*$. This can be conveniently done using an interpolation formula $\tau^{\rm dis}_i(\mu,T)= m_i (\mu,T) v_0^2\hbar/\gamma^2$
which links the ensemble-averaged scattering time (\ref{eq:tau_ensem_ave}) and the effective mass (\ref{eq:eff_mass}) in the entire range of $T$ and $\mu$ of interest.

We find the slope of $\rho_{xy}$ at the DP by expanding Eq.(\ref{eq:rho+i*rho}) in small $n=n_e-n_h$ (see Appendix D for full treatment). The result, which simplifies in each of the regimes $T\gtrsim T_*$ and $T\lesssim T_*$, can be described by a single interpolation formula as
\be\label{eq:lambda}
 \left. \frac{\partial\rho_{xy}}{\partial n} \right |_{n=0}= \frac{\hbar^2 v_0^2}{{\rm min\,}(T_*^2, \pi T^2 /3)}
\,\frac{B}{n_T ec}
,
%
\ee
where only terms first-order in $B$ have been retained.

The SHE coefficient, Eq.(\ref{eq:SHE_DP}), found by combining the results (\ref{eq:lambda}) and (\ref{eq:rho_xx_DP}), and using thermally broadened density of states at the DP derived in Appendix B, $\partial n/\partial \mu=\frac{2\ln 2}{\pi} \frac{k_{\rm B}T}{\hbar^2 v_0^2}$, is
%
\be\label{eq:SHE3}
\left. \theta_{\rm SH} \right|_{n=0}
=
\frac{\lambda E_0^2E_Z}{2 \gamma^2 k_{\rm B}T}
,\quad
E_0= v_0\sqrt{2\hbar eB/c}
,
\ee
where $E_0$ is the cyclotron energy.
The functional form is the same in both regimes,
$\theta_{\rm SH}\propto B^2/T$, with different prefactors $\lambda_{T\gtrsim T_*}=24\ln 2/\pi^2$ and $\lambda_{T\lesssim T_*}=12\ln 2/\pi^2$.
The $1/T$ growth of $\theta_{\rm SH}$ saturates at $k_{\rm B}T\approx \gamma$, reaching maximum value $\theta_{\rm SH, max}  \approx   \frac12\lambda E_0^2E_Z/\gamma^3$.




We expect suspended graphene~\cite{Bolotin08,Du08} to feature an even stronger SHE than graphene on BN. Using typical mobility $\mu_*=2\cdot 10^5{\rm cm^2/V s}$~\cite{Bolotin08}, we estimate $\gamma \sim 65\, {\rm K}$, whereas the temperature dependence of the conductivity at the DP~\cite{Bolotin08} yields $\gamma\sim 10\, {\rm K}$. For either value of $\gamma$, Eq.(\ref{eq:SHE3}) predicts very large values $\theta_{\rm SH}$ at the DP. 


Based on these estimates, we expect strong SHE response already at moderate fields $B\lesssim 1\,{\rm T}$. Besides spin accumulation and locally tunable SHE response, which was discussed above, SHE can also manifest itself in a non-zero Hall voltage in response to spin-polarized currents injected from magnetic contacts.


Since our SHE mechanism does not rely on the relativistic dispersion of excitations, it can also be realized in other zero-gap semiconductors, in particular graphene bilayer. It also applies, with suitable modifications, to the valley degrees of freedom in graphene. It was predicted that a (non-quantizing) magnetic field can lift valley degeneracy and produce a Zeeman-like valley splitting~\cite{Lukyanchuk08}.
This will imbalance the Hall resistivities and result in a valley-Hall effect of a magnitude similar to the SHE.

We thank D. Goldhaber-Gordon, L. M. K. Vandersypen and M. Soljacic for useful discussions and acknowledge support from Naval Research Grant N00014-09-1-0724 (LL).




\section{Appendix A: Spin accumulation}


Here we analyze spin accumulation in the ZSHE regime. Spin accumulation in two-dimensional electron gasses, arising due to weak SHE induced by spin-orbit interaction, was considered in several papers, in particular: S. Zhang, Phys. Rev. Lett. {\bf 85}, 393 (2000). There are several new aspects in our problem that warrant special treatment. Specifically: (i) In the presence of external magnetic field charge transport is described by $\rho_{xx}$ and $\rho_{xy}$ separetly for each spin projection; (ii) Unlike the semiconductor case, the expected SHE response is not necessarily small, $\theta_{\rm SH}\sim 1$.

We consider spin accumulation in the strip geometry, $-w/2<y<w/2$, with current driven along the strip, such that external electric field $\cal E$ is applied along $x$ direction. To solve the transport problem,
we introduce electrochemical potentials for each spin projection, $\varphi_{\uparrow(\downarrow)}=\phi+ n_{\uparrow(\downarrow)}/\nu_{\uparrow(\downarrow)}$,
where $\phi$ is the electric potential, $n_{\uparrow (\downarrow)}$ is the deviation of the local density for spin up (down) from its equilibrium value, and $\nu_{\uparrow(\downarrow)}$ is the density of states for the two spin projections. The equations for the current density are given by,
\be\label{eq:1}
{\bf j}_{\uparrow}({\bf r})=\hat{\sigma}^{\uparrow}\nabla \varphi_\uparrow({\bf r})
, \quad
{\bf j}_{\downarrow}({\bf r})=\hat{\sigma}^{\downarrow}\nabla \varphi_\downarrow({\bf r}),
\ee
where $\uparrow$ and $\downarrow$ label carriers with up-spin and down-spin. Here $\hat{\sigma}^{\uparrow}$ and $\hat{\sigma}^{\downarrow}$ are $2\times 2$ matrices describing the longitudinal and Hall conductivity. The continuity equation for each spin projection can be written as
\be\label{eq:2}
\nabla {\bf j}_\uparrow ({\bf r})=-\gamma_s( n_\uparrow({\bf r})- n_\downarrow({\bf r}))
, \quad
\nabla {\bf j}_\downarrow ({\bf r})=-\gamma_s(n_\downarrow({\bf r})- n_\uparrow({\bf r})),
\ee
where $\gamma_s$ is the rate of spin relaxation. The terms with time derivative are omitted, as appropriate for the DC transport.

The transport equations should be supplemented by the condition of electro-neutrality, $ n_\uparrow({\bf r})=- n_\downarrow({\bf r})$. We solve transport equations in a strip geometry, where current is driven by an electric field parallel to the strip, with the boundary conditions of zero normal current at the strip edges, $j_y(y=\pm \frac12 w)=0$. We obtain the spin density $n_s=n_{\uparrow}-n_{\downarrow}$ profile across the strip
\be\label{eq:spin_d}
n_s(y)= {e{\cal E}  \bar\nu\ell_s \theta_{\rm SH}} \frac{\sinh \frac{y}{\ell_s} }{\cosh \frac{ w}{2 \ell_s}}
,\quad
\theta_{\rm SH}=\frac{\sigma_{xy}^\uparrow}{\sigma_{xx}^\uparrow}-\frac{\sigma_{xy}^\downarrow}{\sigma_{xx}^\downarrow}
,
\ee
where $\bar\nu=2/(\nu_\uparrow^{-1}+\nu_\downarrow^{-1})$ is the average density of states for the two spin projections, $\cal E$ is the electric field, $\ell_s$ is the spin relaxation length
\be\label{eq:ls}
\ell_s^2=\frac{ \bar\sigma_{xx}}{2\bar\nu \gamma_s}
,\quad
\frac{1}{\bar\sigma_{xx}}=\frac{1}{2}\left( \frac{1}{\sigma_{xx}^\uparrow}+  \frac{1}{\sigma_{xx}^\downarrow} \right)
.
\ee
For a narrow strip, $w\ll \ell_s$, spin relaxation can be ignored. In this case, the expression for $n_s(y)$ agrees with the estimate (\ref{eq:SHE_DP}) in the main text.


%

To estimate the numerical value of the spin density at the DP, where it is maximum, we make several assumptions. First, we assume that
the density of states is disorder-broadened at the Dirac point, $\bar\nu =\nu_\uparrow=\nu_\downarrow\approx \frac{\gamma}{\pi \hbar^2 v_0^2}$, where $\gamma$ is the disorder strength parameter defined in Eq.(\ref{eq:scat_time}) of the main text. For an estimate we use the value $\gamma\sim 100\, {\rm K}$, which corresponds to density inhomogeneity $\Delta n \approx 2 \nu(0)\gamma\approx 10^{10} {\rm cm^{-2}}$ typical for graphene on BN substrate. Second, we will assume $w=2 {\rm \mu m}$, and
for the SHE coefficient, we will assume $\theta_{\rm SH}=0.1$, as estimated in the main text.
Taking ${\cal E}=10\, {\rm V/\mu m}$, as in Ref.~\cite{current_density}, for the spin density at the edge we find
\be\label{eq:spin_d_size}
n_s(y=\pm w/2)\approx 3\cdot 10^{10} \,\, {\rm cm^{-2}}.
\ee
This is at least five orders of magnitude the spin density per atomic layer observed in GaAs~\cite{Kato04}, $n_s\approx 5\times 10^5 \, {\rm cm^{-2}}$.
Such giant spin accumulation is due to the larger value of $\theta_{\rm SH}$ in graphene, and due to the fact that graphene can sustain large current densities and electric fields.

\section{Appendix B: The carrier density and the density of states}

We start with deriving a closed form expression for the carrier density, evaluated separately for the electrons and holes with a fixed spin projection:
$n_i(\mu,T)= 2\int \frac{d^2\vec{p}}{(2\pi)^2}  f_i(\vec{p},\mu,T)$, where the distribution function $f_i$ is defined by Eq.(\ref{eq:distributionE}) of the main text with ${\bf a}_i=0$. Integrating over $p$, we find the dependence of $n_{e(h)}$ on temperature and chemical potential as power series in $g=e^{\mu/k_{\rm B}T}$
\bea\nonumber
n_{e(h)}(\mu,T) &=& \frac{1}{\pi}\lp \frac{k_{\rm B}T}{\hbar v_0}\rp^2\lp g-\frac{g^2}{2^2}+\frac{g^3}{3^2}-\frac{g^4}{4^2}+...\rp
\\
&=& \frac{1}{\pi}\lp \frac{k_{\rm B}T}{\hbar v_0}\rp^2
\int_0^g\frac{\ln(1+g')}{g'}dg'
.
\eea
Although the series converge only for $-1< g\le 1$, the final expression is valid for both $g<1$ and $g>1$. 
%
At the DP, setting $\mu=0$, we obtain
\be
n_{e(h)}(\mu=0)=\frac{\zeta(2)}{2\pi}\lp \frac{k_{\rm B}T}{\hbar v_0}\rp^2
=\frac{\pi}{12}\lp \frac{k_{\rm B}T}{\hbar v_0}\rp^2
.
\ee
This quantity, denoted as $n_T$ in Eq.(\ref{eq:rho_xx_DP}) of the main text and elsewhere, gives the density of thermally excited carriers (electrons or holes) with a fixed spin projection.

Next, we analyze the density of states $\nu=\partial n/\partial\mu$.
The density of states per one spin projection is represented by a sum of an electron and a hole contribution
%
\be
\nu_{\uparrow(\downarrow)} =\frac{\partial n_e}{\partial \mu}
- \frac{\partial n_h}{\partial \mu}
,
\ee
where the minus sign reflects the fact that electrons and holes have opposite chemical potentials, $\mu_e=-\mu_h=\mu$. 
%
We use the above expression for $n_{e(h)}(\mu)$ to find
%
\bea
\nu_{\uparrow(\downarrow)} &=&\lp \ln(1+g)+\ln(1+g^{-1})\rp
\frac{k_{\rm B}T}{\pi\hbar^2 v_0^2}
\\
&=&\ln\lp 2\cosh\frac{\mu}{2k_{\rm B}T}\rp
\frac{2k_{\rm B}T}{\pi\lp \hbar v_0\rp^2}
.
\eea
This expression describes temperature-broadened density of states:
\be\label{eq:nu_cases}
\nu^{\uparrow(\downarrow)}_{|\mu|\gg k_{\rm B}T}=\frac{|\mu|}{\pi\lp \hbar v_0\rp^2}
,\quad
\nu^{\uparrow(\downarrow)}_{|\mu|\ll k_{\rm B}T}=\frac{{2\ln 2}}{\pi} \frac{k_{\rm B}T}{\lp \hbar v_0\rp^2}
\ee
with a crossover temperature $k_{\rm B}T=|\mu|/(2\ln 2)$.


\section{Appendix C: The effect of drag between carriers of opposite spin}

Our analysis of transport coefficients in the main text was based on a simplified model which neglected drag between carriers of opposite spin. To evaluate the accuracy of this approach, here we discuss a more general model which accounts for drag between carriers of either spin. To emphasize this difference, we will use notation $\tilde\eta$ for the drag coefficient instead of $\eta$ used in the main text. As we will see, the more general approach predicts an essentially identical behavior of transport coefficients, and a qualitatively similar behavior of the SHE coefficient.

We consider coupled transport of four carrier species.
For simplicity, we will take the drag coefficient values to be the same for all carrier species, a reasonable approximation at high $k_{\rm B}T$. Ensemble-averaged velocities and momenta in the presence of electric and magnetic fields are governed by Eq.(\ref{eq:electron}) of the main text, which we duplicate here for reader's convenience,
\be\label{eq:4comp}
q_i \lp {\bf E}+\frac{\vec V_i}{c}  \times {\bf B}\rp=-\frac{\vec P_i}{\tau^{\rm dis}_{i}}-\tilde\eta\sum_{i'} n_{i'}   (\vec V_i-{\vec V}_{i'}).
\ee
The quantities $n_i$, $m_i$, $\tau_i^{\rm dis}$ are defined as in the main text, with the chemical potential $\mu$ replaced by $\mu\pm E_Z/2$ for spin up (down) electrons. The value of the drag coefficient is found by matching analytic results\cite{Kashuba08,Mueller09}. Because Eqs.(\ref{eq:4comp}) account for drag between all carrier species, not just those with parallel spins, the value $\tilde\eta$ found below is different from the one obtained in the main text.

The transport coefficients are determined as follows. First, for given values $\mu$ and $T$, we obtain $n_i$, $m_i$ by numerically evaluating integrals in the relation 
\be
n_i(\mu,T)= 2\int \frac{d^2\vec{p}}{(2\pi)^2}  f_i(\vec{p},\mu,T)
,
\ee 
and in Eq.(\ref{eq:eff_mass}). For ensemble-averaged scattering time $\tau_{i}^{\rm dis}$, we use the interpolation formula $\tau^{\rm dis}_i(\mu,T)= m_i (\mu,T) v_0^2\hbar/\gamma^2$
which links the ensemble-averaged scattering time (\ref{eq:tau_ensem_ave}) and the effective mass (\ref{eq:eff_mass}) in the entire range of $T$ and $\mu$ of interest.

After that, we solve the four coupled linear equations (\ref{eq:4comp}) and find the currents of the spin-up and spin-down electrons, 
\be
{\bf j}_{\uparrow}=n_{e\uparrow } e {\bf V}_{e\uparrow }-n_{h\uparrow} e{\bf V}_{h\uparrow }
,\quad
{\bf j}_{\downarrow}=n_{e\downarrow } e {\bf V}_{e\downarrow }-n_{h\downarrow} e{\bf V}_{h\downarrow }. 
\ee
Then the charge and spin currents are expressed through ${\bf j}_{\uparrow}$ and ${\bf j}_{\downarrow}$ as follows,
\be\label{eq:charge/spin_current}
{\bf j}_{\rm c}={\bf j}_\uparrow+{\bf j}_\downarrow
,\quad 
{\bf j}_{\rm s}= {\bf j}_{\uparrow}-{\bf j}_{\downarrow},
\ee
These expressions can be used to calculate the charge and spin conductivity tensors. We obtain the dimensionless SHE coefficient by evaluating the ratio of the transverse spin current and the longitudinal charge current, 
\be\label{eq:xi_SH}
\xi_{\rm SH}=2 j_{s, \perp} /j_{c, \parallel }
\ee
where the transverse and longitudinal components are taken with respect to the electric field ${\bf E}$.


\begin{figure}
\includegraphics[width=3.3in]{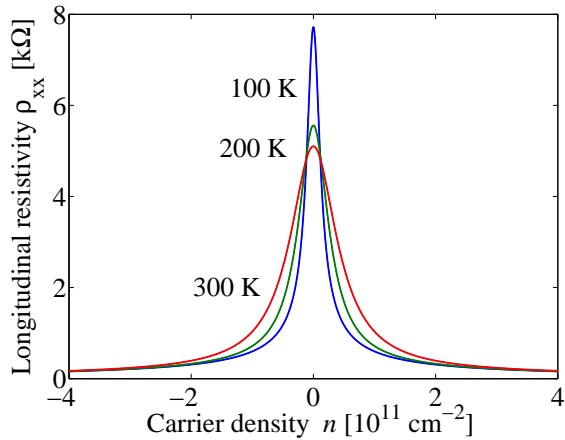}
\caption[]{
The longitudinal resistivity, obtained from the model (\ref{eq:4comp}) which takes into account drag between carriers of either polarity and spin, shown for several temperatures. Parameters used: $B=0\,{\rm T}$, $\gamma=100\, {\rm K}$, $\tilde\eta=1.15\hbar$. The peak in $\rho_{xx}$ gets broadened as the temperature increases, as $\rho_{xx}(0)$ approaches the high-temperature limiting value, Eq.(\ref{eq:rho_xx_limit}).}
\label{fig7}
\end{figure}

For charge transport, this model leads to the behavior of transport coefficients which is essentially identical to that obtained from the simplified model used in the main text, Eq.(\ref{eq:rho+i*rho}),
albeit with a doubled value of the drag coefficient, $\eta\to 2\tilde\eta$. To see this, we note that the forces $q_{e(h)}\lp \vec E+\frac1{c}\vec v\times\vec B\rp$ that drive transport are the same for the up-spin and down-spin carriers. Further, the quantities $n_i$, $m_i$, $\tau^{\rm dis}_i$ coincide for spin up and down when $E_z=0$. Therefore,  in the limit $E_Z\to 0$, the relations (\ref{eq:4comp}) can be satisfied by $\vec v_e^\uparrow= \vec v_h^\downarrow$, $\vec v_h^\uparrow= \vec v_e^\downarrow$. Taking this into account and eliminating variables for one spin projection we obtain equations for the other projection which are identical to the equations in the main text up to a replacement $\eta \to 2\tilde\eta$. Small but finite $E_Z$ changes the transport coefficients, however the difference between the spin-up and spin-down remains small as long as $E_Z\ll k_{\rm B}T,\gamma$. 
 
Following the same reasoning as in the main text, we fix the drag coefficient value at the half the drag coefficient used in the simplified model: $2\tilde\eta=2.3\hbar$. We obtain the density dependence $\rho_{xy}(n)$ which is very similar to that shown in Fig.~\ref{fig2} of the main text. Behavior of the longitudinal resistivity $\rho_{xx}(n)$ at $B=0$ is illustrated in Fig.~\ref{fig7}. The peak in $\rho_{xx}$ at the DP becomes lower and broader as the temperature is increased. The resistivity at the DP decreases, at high temperatures $k_{\rm B}T\gg \gamma$ saturating at the value 
\be\label{eq:rho_xx_limit}
\rho_{xx}=\frac12\rho_{xx}^{\uparrow(\downarrow)}=\tilde\eta /e^2\approx 5 {\rm k\Omega}
,
\ee
where the factor $1/2$ is introduced to convert the resistivity for one spin projection, given by Eq.(\ref{eq:rho_xx_DP}) of the main text, to net resistivity. The results in Fig.~\ref{fig7} show that $\rho_{xx}(0)$ is very close to this limiting value already at $T=300\, {\rm K}$, indicating that transport at these temperatures is dominated by electron-hole collisions.

\begin{figure}
\includegraphics[width=3.3in]{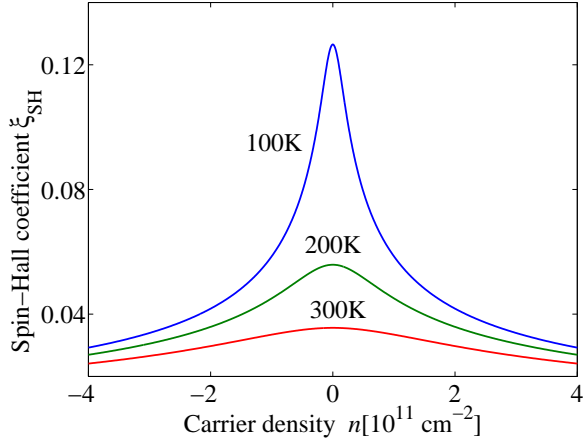}
\caption[]{
The SHE coefficient $\xi_{\rm SH}$, Eq.(\ref{eq:xi_SH}), obtained from the model (\ref{eq:4comp}) which takes into account drag between carriers of either polarity with parallel an opposite spins. Parameters used: $B=1\,{\rm T}$, $\gamma=100\, {\rm K}$, $\tilde\eta=1.15\hbar$. Similarly to $\theta_{\rm SH}$, illustrated in Fig.~\ref{fig1}, $\xi_{\rm SH}$ exhibits a sharp peak at the DP, which gets more pronounced as $T$ is lowered. The quantities $\xi_{\rm SH}$ and $\theta_{\rm SH}$ are identical at low $T$, when transport is dominated by scattering on disorder. At high $T$, when transport is dominated by the effects of electron-hole drag, the peak in $\xi_{\rm SH}$ is significantly broader than the peak in $\theta_{\rm SH}$.}
\label{fig6}
\end{figure}


The behavior of the SHE coefficient $\xi_{\rm SH}$, Eq.(\ref{eq:xi_SH}), shown in Fig.\ref{fig6}, is overall similar although not identical to the behavior of $\theta_{\rm SH}$ found in the main text. Near the DP, $\xi_{\rm SH}$ reaches large values, similar to those of $\theta_{\rm SH}$ (see Fig.~\ref{fig1} in the main text). The peak in $\xi_{\rm SH}$ is, however, significantly broader. Furthermore, $\xi_{\rm SH}$ exhibits a monotonic decay away from the DP; in contrast, $\theta_{\rm SH}$ exhibits a sign change. We believe the sign change to be peculiar for the two-component model which ignores the drag between spin-up and spin-down carriers.

The SHE coefficient $\xi_{SH}$ determines physical observables, such as spin accumulation density, which is given by Eq.(\ref{eq:SHE_DP}) of the main text with a substitution $\theta_{\rm SH}\to \xi_{\rm SH}$. The two models therefore predict the same values of $n_s$ at the DP.

\section{Appendix D: Analytic estimates for the SHE coefficient}

Here we derive the expression for $\partial\rho_{xy}/\partial n$, Eq.(\ref{eq:lambda}), and an estimate for the SHE coefficient, Eq.(\ref{eq:SHE3}) of the main text. First, we consider the high-temperature regime, $T\gg T_*$, where transport coefficients are dominated by the effects of electron-hole drag. In this limit, the first term in the numerator of Eq.(\ref{eq:rho+i*rho}), which is small compared to the other two terms, can be neglected. Furthermore, the last term in the denominator, which is quadratic in density, cannot affect the slope of $\rho_{xy}$ at the DP, and we can drop it as well. Then Eq.(\ref{eq:rho+i*rho}) is simplified as
\be\label{eq:rho+i*rho-2}
\lp \rho_{xx}+i\rho_{xy}\rp^{\uparrow(\downarrow)} \approx \frac{\eta}{e^2}\frac{\frac{n_e}{m_h}\tilde\gamma_e + \frac{n_h}{m_e}\tilde\gamma_h}{\frac{n_e}{m_e}\tilde\gamma_h +\frac{n_h}{m_h}\tilde\gamma_e}
.
\ee
Using the relations $\tilde\gamma_i=\frac{1}{\tau_i^{\rm dis}}-i\Omega_i$ and the interpolation formula $\tau_i^{\rm dis}=m_i v_0^2\hbar/\gamma^2$, after simple algebra we obtain 
\be\label{eq:rho+i*rho-3}
\lp \rho_{xx}+i\rho_{xy}\rp^{\uparrow(\downarrow)}\approx \frac{\eta}{e^2}\frac{\gamma^2(n_e+n_h)/v_0^2\hbar-ieBn/c}{\gamma^2(n_e+n_h)/v_0^2\hbar+ieBn/c}
,
\ee
where $n=n_e-n_h$ is the carrier density. Then, using the fact that $n_e+n_h=2n_T+O(n^2)$, and expanding to linear order in $B$, we obtain 
\be
\rho_{xx}=\frac{\eta}{e^2}
,\quad
\rho_{xy}= \frac{\hbar^2 v_0^2}{T_*^2}\frac{B}{n_T ec} n
.
\ee
The second relation gives the slope of the Hall resistivity at the DP, Eq.(\ref{eq:lambda}) of the main text (case $T>T_*$). 

In the low-temperature regime, $T\ll T_*$, we can neglect the second and third terms in the numerator of Eq.(\ref{eq:rho+i*rho}),as well as the third term in the denominator, which gives 
\be\label{eq:rho+i*rho-4}
\lp \rho_{xx}+i\rho_{xy}\rp^{\uparrow(\downarrow)}\approx \frac{1}{e^2}\frac{\tilde\gamma_e \tilde\gamma_h}{\frac{n_e}{m_e}\tilde\gamma_h +\frac{n_h}{m_h}\tilde\gamma_e}
.
\ee
Once again using the relations $\tilde\gamma_i=\frac{1}{\tau_i^{\rm dis}}-i\Omega_i$ and the interpolation formula $\tau_i^{\rm dis}=m_i v_0^2\hbar/\gamma^2$, we rewrite the above expression as 
\be\label{eq:rho+i*rho-4b}
\lp \rho_{xx}+i\rho_{xy}\rp^{\uparrow(\downarrow)}\approx \frac{1}{e^2}\frac{\gamma^4/\hbar^2 v_0^4}{\gamma^2  (n_e+n_h)/\hbar v_0^2+ieB n  /c}
,
\ee
where we have neglected the term proportional to $\Omega_e \Omega_h$ in the numerator, which is of the order $B^2$. 
Expanding to linear order in $n$, we obtain 
\be
\rho_{xx}= \frac{\gamma^2}{2n_Te^2 \hbar v_0^2}
,\quad
\rho_{xy}=\frac{B }{n_T e c} \frac{n}{4 n_T}.
\ee
Using the relation $n_T=\frac{\pi}{12}\left( \frac{k_B T}{\hbar v_0} \right)^2$, we obtain the slope of $\rho_{xy}$ at the DP, Eq.(\ref{eq:lambda}) of the main text (case $T<T_*$).




Finally, we obtain the SHE coefficient at the DP, defined in Eq.(\ref{eq:SHE_DP}) of the main text. This is done by combining the results for $\p\rho_{xy}/\p n$, which we have just derived, with the result for thermally broadened density of states at the DP, Eq.(\ref{eq:nu_cases}),
and the above results for $\rho_{xx}$ at the DP. This gives Eq.(\ref{eq:SHE3}) of the main text.


\begin{references}
\bibitem{HDDheads} Current HDD
heads can generate a very strong and spatially localized magnetic field, reaching several
thousands of Gauss (close to 1
Tesla) in areas around $100{\rm nm}\times 35{\rm nm}$.
\end{references}
\end{document}